\documentclass[twocolumn,prl,aps]{revtex4-1}
\usepackage{amsmath}
\usepackage{sistyle}

\usepackage{graphicx}%
\usepackage{dcolumn}%
\usepackage{bm}%

\begin{document}
\title{Heralded amplification of photonic qubits}

\author{Natalia Bruno}
\address{Group of Applied Physics, University of Geneva, CH-1211 Geneva 4, Switzerland}
\author{Vittorio Pini}
\address{Group of Applied Physics, University of Geneva, CH-1211 Geneva 4, Switzerland}
\author{Anthony Martin}
\address{Group of Applied Physics, University of Geneva, CH-1211 Geneva 4, Switzerland}
\author{Varun Verma}
\address{National Institute of Standards and Technology, 325 Broadway, Boulder, Colorado 80305, USA }
\author{Sae Woo Nam}
\address{National Institute of Standards and Technology, 325 Broadway, Boulder, Colorado 80305, USA }
\author{Richard Mirin}
\address{National Institute of Standards and Technology, 325 Broadway, Boulder, Colorado 80305, USA }
\author{Adriana Lita}
\address{National Institute of Standards and Technology, 325 Broadway, Boulder, Colorado 80305, USA }
\author{Francesco Marsili}
\address{et Propulsion Laboratory, California Institute of Technology, 4800 Oak Grove Dr., Pasadena, California 91109, USA }
\author{Boris Korzh}
\address{Group of Applied Physics, University of Geneva, CH-1211 Geneva 4, Switzerland}
\author{F\'{e}lix Bussi\`{e}res}
\address{Group of Applied Physics, University of Geneva, CH-1211 Geneva 4, Switzerland}
\author{Nicolas Sangouard}
\address{Department of Physics, University of Basel, CH-4056 Basel, Switzerland}
\author{Hugo Zbinden}
\address{Group of Applied Physics, University of Geneva, CH-1211 Geneva 4, Switzerland}
\author{Nicolas Gisin}
\address{Group of Applied Physics, University of Geneva, CH-1211 Geneva 4, Switzerland}
\author{Robert Thew}
\email{robert.thew@unige.ch}
\address{Group of Applied Physics, University of Geneva, CH-1211 Geneva 4, Switzerland}

\begin{abstract}
We demonstrate heralded qubit amplification for Time-Bin and Fock-state qubits in an all-fibre, telecom-wavelength, scheme that highlights the simplicity, the stability and potential for fully integrated photonic solutions. 
Exploiting high-efficiency superconducting detectors, the gain, the fidelity and the performance of the amplifier are studied as a function of loss. We also demonstrate the first heralded Fock-state qubit amplifier without post-selection. This provides a significant advance towards demonstrating Device-Independent Quantum Key Distribution as well as fundamental tests of quantum mechanics over extended distances.
\end{abstract}

\maketitle

\section{Introduction}

Terrestrial quantum communication mainly relies on the efficient transmission of photonic states through fibre optic networks~\cite{Gisin2007}. 
Photon loss represents one of the most significant challenges for photonic quantum technologies to overcome not only for applications in quantum communication, but for metrology~\cite{Giovannetti2011}  and fundamental tests of quantum physics~\cite{Brunner2014}. 
%
%
Recently, a novel approach for QKD has been proposed. It relies on a Bell test and makes the security device independent (DI-QKD), i.e. the measurement devices and the sources can be treated as black boxes \cite{Acin2007,Mayers:1998aa,Gisin2010}. 
However, this requires the violation of a Bell inequality free of the detection loophole\cite{Brunner2014,Christensen2013,Giustina2013}.
Heralded Qubit Amplification (HQA)~\cite{T.C.Ralph2009,Gisin2010,Pitkanen2011,Curty:2011aa} provides a teleportation based solution to overcome transmission loss, e.~g. in the fiber optics used for telecommunication, thus providing a route towards a loophole-free Bell test over tens of kilometers and eventually DI-QKD.

Heralded amplification has previously been demonstrated in proof-of-principle experiments in the visible~\cite{Kocsis2013} and telecom regimes~\cite{Osorio2012,Bruno2013}. 
There has also been interest in exploiting this for continuous variable systems to reduce noise~\cite{Ferreyrol2010,Barbieri2011,Neergaard-Nielsen:2013aa} and extend transmission distances~\cite{Chrzanowski2014,Fuwa2014,Alexander-E.-Ulanov:2015aa}. 
Fundamental tests of single photon addition and subtraction have also been explored in the same context~\cite{Fiur2009,ZavattaA.2011}. 
Here, we work in the telecom regime and develop simple, all-fibre, linear optic circuits for schemes adapted for quantum communication; exploiting Time-bin~\cite{Brendel1999} and Fock-state~\cite{Sangouard2007} encodings. In the case of the Fock-state scheme we demonstrate the first HQA without post-selection.

\section{Principle}

The principle for heralded qubit amplification follows directly from a teleportation protocol.

\begin{figure}
\includegraphics[width=\columnwidth]{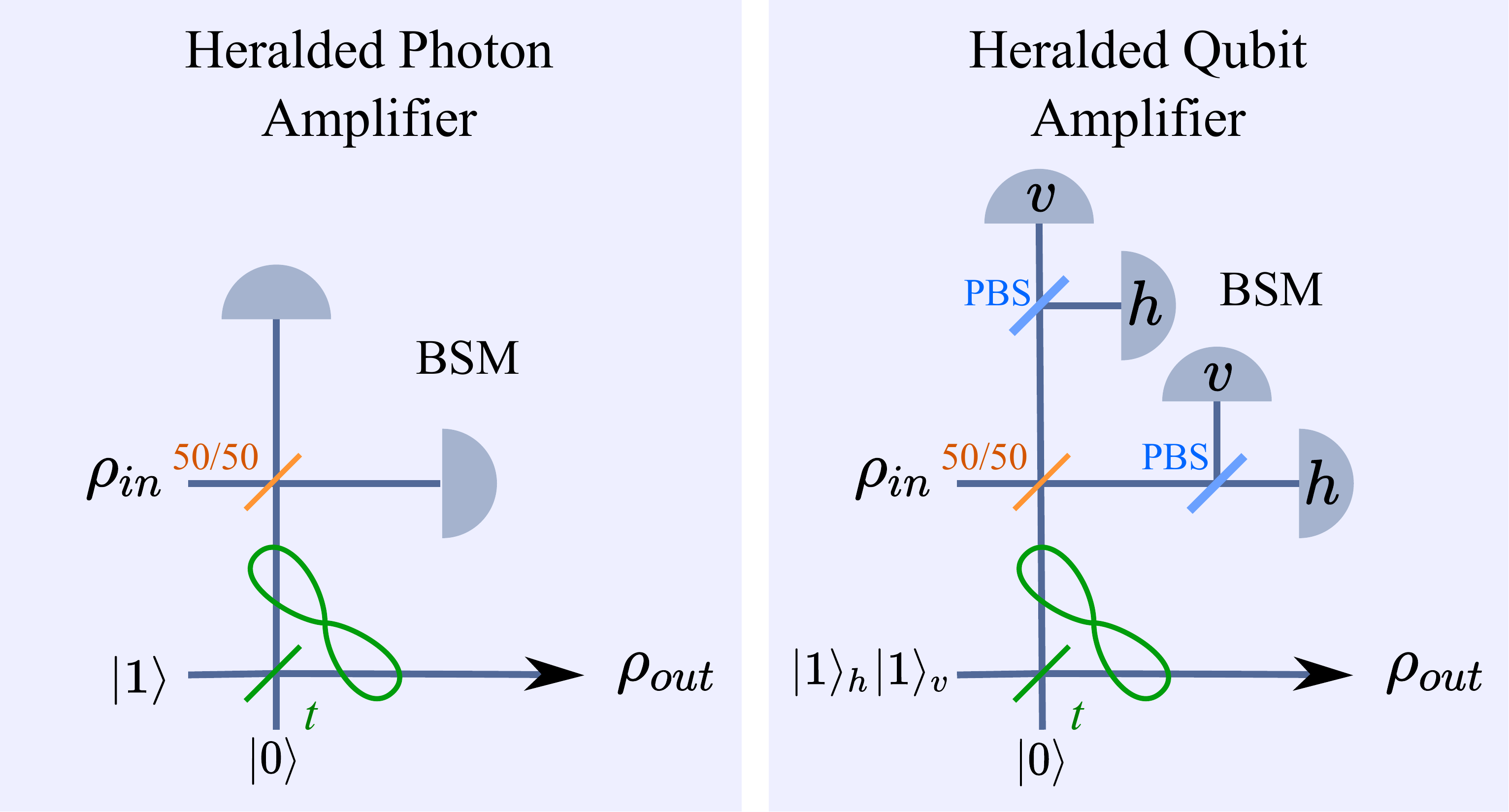}
\caption{Left: heralded single photon amplifier. Right: linear circuit for the heralded polarisation-qubit amplifier.}\label{fig1}
\end{figure}

 Consider \figurename{~\ref{fig1} }, where an unknown input qubit state is combined with one mode of an entangled state.

An initial qubit state $|\psi_{\text{in}} \rangle = \alpha|\phi\rangle + \beta|\phi^{\perp}\rangle $, is transmitted, for example, through a fibre optic channel. It will undergo loss, such that the resulting state is in a statistical mixture of the initial state and vacuum:
\begin{equation}
\rho_{\text{in}} = (1-P_{\text{in}}) \vert 0 \rangle \langle 0 \vert + P_{\text{in}} \vert \psi_{\text{in}} \rangle \langle \psi_{\text{in}} \vert,
 \end{equation}  
where $P_{\text{in}}$ is the channel transmission. This will be the input state for the HQA, which also requires mutually orthogonal auxiliary states that span the relevant qubit space, $|\psi\rangle_a \equiv |\phi \rangle \otimes | \phi^{\perp} \rangle $. In the case of Fock-state qubits, as shown in \figurename{~\ref{fig1} (Left)}, the ancilla state is  $|1\rangle\otimes|0\rangle$.
This is easily extended to polarisation qubits, or similarly time-bin qubits, by coherently combining two single-photon amplifiers - one for each mode, see~\figurename{~\ref{fig1} (Right)}.

 A Bell state measurement (BSM) acting on the initial qubit and an ancilla qubit teleports the input state into the output one, which is the other mode of the entangled state. The use of an unbalanced beam splitter (with transmission $t$) allows for the teleportation to be biased so as to reduce the vacuum contribution.
The resulting output state is still a mixture between the vacuum and the initial qubit, up to a unitary transformation $U$ that depends on the result of the BSM. However, it now has different relative weights:
\begin{equation}
\rho_{out}  =  (1 - G(t) P_\text{in})\vert 0 \rangle \langle 0 \vert + G(t) P_\text{in} U \vert \psi_{\text{in}} \rangle \langle \psi_{\text{in}} \vert U^{-1}.\label{eq:rhoout}
\end{equation}
The gain $G(t)$ is the ratio between the probability of having a photon in the output given a succesful BSM, $P_\text{out}$, and the probability of having a qubit at the input, $P_\text{in}$.
In the ideal case, it tends to $t/(1-t)$ in the limit of high loss. In practice, considering realistic experimental conditions, it also depends on the coupling efficiency of the ancilla $P_{\text{a}}$ and the detection efficiency $\eta$. 
Moreover, having non photon-number-resolving (PNR) detectors increases the probability of heralding the vacuum, and therefore it reduces the gain with respect to the ideal case. 
 If one takes into account non PNR detectors with efficiency $\eta$ on the BSM and a probability $P_\text{a}$ of having an ancilla (for simplicity, we consider $P_\text{a}$ to be the same for both auxiliary photons), we can define the gain as:
\begin{equation}\label{G}
G(t) = \frac{P_\text{out} }{P_\text{in}} = \frac{P_\text{a} t}{P_\text{a} (1-t)(1-P_\text{in} \eta) +P_\text{in}}.
\end{equation}
As we can see, a low ancilla input probability $P_\text{a}$ and a poor detection efficiency $\eta$ reduce the gain.
Notice that a proposed modification of the original scheme allows one to improve the gain by eliminating the case in which two auxiliary photons are reflected~\cite{Pitkanen2011}. However, this scheme is more demanding from an experimental point of view as it requires a phase-stabilised interferometer between the unbalanced beam splitter and the BSM. 

\section{Experimental setup}
\subsection{Concept}
 In this work, we use time-bin qubits, e.g. $\vert \psi_\text{in} \rangle = \frac{1}{\sqrt{2}}\left( \vert s_H \rangle + e^{i \Delta \phi} \vert \ell_V \rangle \right) $.
 The polarisation is used to label and switch the {\it short} and {\it long} paths to improve efficiency~\cite{Townsend1993,Thew2006}, requiring the ancilla state to be of the form $|\psi\rangle_a = |s_H\rangle\otimes|\ell_V \rangle $.

\begin{figure*}
\includegraphics[width=1.9\columnwidth]{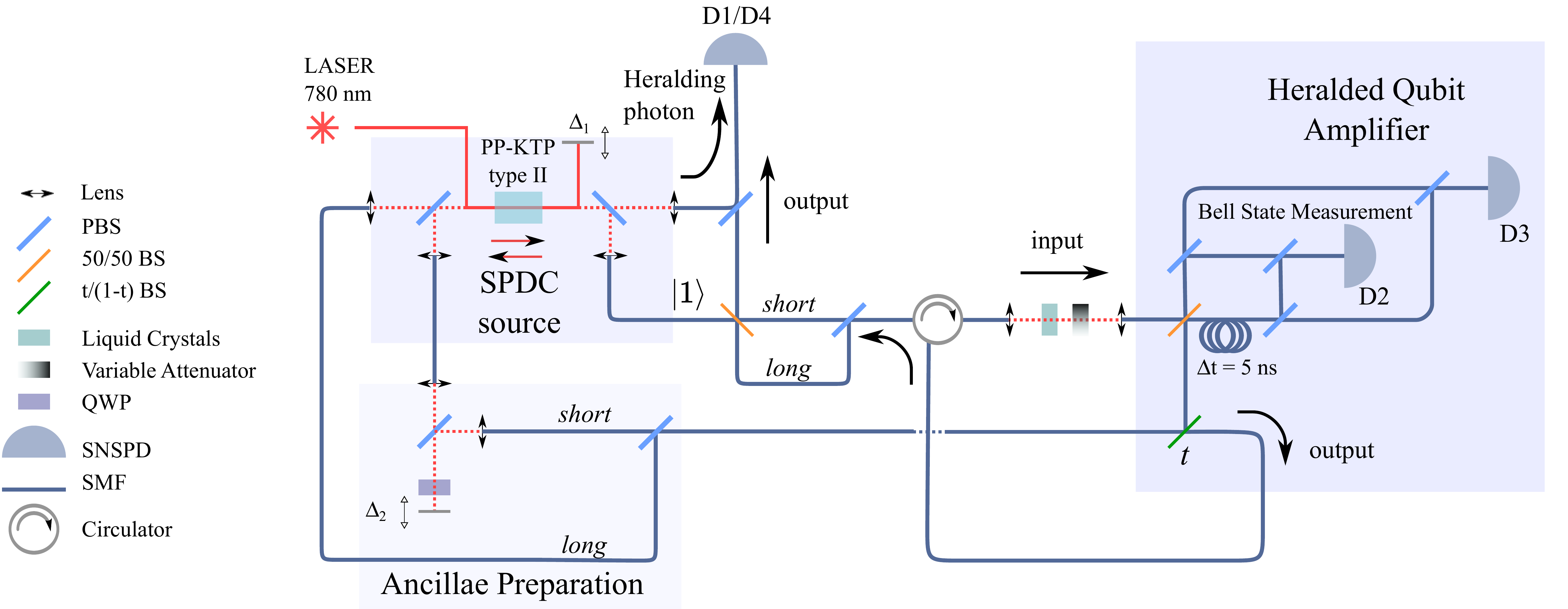}
\caption{Scheme of the time-bin qubit amplifier. A heralded single photon is prepared in a time-bin qubit and subsequently sent through a lossy channel. The amplification is heralded by a 3-fold coincidence given by  a click in the detector D1, which heralds the input photon, followed by D2 and D3, heralding its teleportation. 4-fold coincidence with the output photon in detector D4 are recorded. We use MoSi SNSPDs with $70\%$ detection efficiency at 1550\,nm, at a temperature T\,$<$\,2.5\,K~\cite{V.-B.-Verma:2015aa}.
 Two polarisation modes which accompany the two temporal modes, \textit{short} and \textit{long}, allow one to separate the $\ell$ and $s$ components in the analysis  without further loss. A temporal delay introduced in one of the outputs of the BSM allows one to project the input modes onto two Bell states ($\Psi^{\pm}$) using two detectors instead of four. The analysis is performed in the same interferometer used for the preparation, with the aid of a fibred circulator. In this way, active stabilisation of the phase is not needed. The relative phase in the input qubit is controlled after the interferometer, taking again advantage of the two polarisation modes, using a variable liquid crystal retarder.}\label{fig2}
\end{figure*}

\figurename{~\ref{fig2}} shows the experimental setup where four indistinguishable photons are generated by Spontaneous Parametric Down Conversion (SPDC). Detector $D_1$ heralds a single photon that is sent into an interferometer to prepare the time-bin qubit. The other source provides the two ancillae photons, where one is delayed with respect to the other by $\Delta_2$, corresponding to the path-length difference in the time-bin interferometer. The delay between the {\it short} photon from each source is determined by $\Delta_1$. It is critical that the arrival times of the two ancilla photons match the two qubit modes for the BSM to occur with high fidelity. The phase of the initial qubit can be varied and loss can be added before the HQA. The output state is then sent back to the same interferometer to verify the fidelity with respect to the initial state. 

We then analyse the result based on the four-fold coincidences between the heralded single photon $D_1$, a successful BSM $D_2$ and $D_3$, and the output state $D_4$ detected by MoSi superconducting nanowire single photon detectors (SNSPD) with  an efficiency of $70\,\%$~\cite{V.-B.-Verma:2015aa}. By exploiting the fast recovery time (80\,ns) of these detectors, each one is used twice in each run of the experiment, i.e. with two different and well distinguished arrival times discriminated via the clock of the pump laser.

\subsection{Heralded single photon source.} The heralded single photon source used in this work is based on a SPDC process in a 3\,cm bulk Periodically Poled Potassium Titanyl Phospate (PP-KTP) crystal, with type II phase matching conditions. With this material, adjusting the poling period, the crystal length and the pump pulse duration, it is possible to achieve quasi phase matching conditions that allow the generation of spectrally separable photon pairs \cite{Grice2001,Mosley2008,Bruno2014}. 
Moreover, purity is optimised when the photons are selected in a single spatial mode with a high heralding efficiency \cite{Guerreiro2013}.

A 76 MHz picosecond pulsed laser at 772 nm pumps the crystal in a double-pass configuration, generating pairs of photons in each direction, for a total of four photons, degenerate at 1544 nm. The setup is symmetrical: for each of the two passes of the pump through the crystal, the two daughter photons are separated by a polarisation beam splitter (PBS) and then coupled into single mode fibres. The achieved heralding  efficiency after coupling into single mode fibers (SMF) is measured to be $0.86\pm0.04 $ for the qubit (pair generated in the first passage through the crystal), $0.80\pm0.04$ on average for the two ancillae (generated in the second passage through the crystal), not considering the detection efficiency.
A Hong-Ou-Mandel (HOM) interference experiment between two photons produced in independent sources gives a visibility of $0.92 \pm 0.03$, indicating high spectral purity without any filtering~\cite{Bruno2014}. To increase the single mode character of the generated photons, two 100\,GHz filters are placed before the heralding, BSM, detectors, allowing one to herald the output state in a well defined spectral mode.

The path of the qubit to the amplifier includes the preparation interferometer, a circulator and a free space path where a liquid crystal retarder is placed, allowing one to vary the relative phase between the two components of the qubit ($s$ and $\ell$, which have orthogonal polarisations, see \figurename{\ref{fig2}}).
The transmission of these elements is $0.55\pm0.04$.
Two HOM interference measurements between the qubit in the input mode and each of the two auxiliary photons in the modes $short$ and $long$ are performed in order to adjust the arrival times onto the $50/50$ beam splitter (BS) needed for the BSM. The total transmission of this part  of the setup is $0.37\pm0.02$. 
 In this experiment the excess loss of the ancillae strongly limits the value of $P_{\text{out}}$. However, this is due to the need of controlling and adjusting the temporal delays of the interfering photons. In principle, this quantity can be considerably improved when fixing all the parameters at their optimal values.

After the $50/50$ BS a system of four polarisation beam splitters (PBS) and a fixed delay of $5 ns$ on one of the BS outputs allow one to project onto the two Bell states $\Psi^{\pm}$, using two detectors (D2 and D3 in \figurename{~\ref{fig2}})  instead of four as shown in the right of \figurename{~\ref{fig1}}. 

\subsection{Detectors.} The MoSi detectors were fabricated to obtain maximum efficiency at a wavelength of 1550~nm and for operation at 2.5~K in a two-stage closed-cycle cryocooler. Their fabrication and characterization is detailed in~\cite{V.-B.-Verma:2015aa}. A gold mirror was fabricated on top of Titanium on a 3 inch Silicon wafer using electron-beam evaporation, and was photolithographically patterned using a lift-off process. A SiO$_2$ space layer between the mirror and the MoSi was then deposited by plasma-enhanced chemical vapour deposition (PECVD). A~6.6 nm-thick Mo$_x$Si$_{1-x}$ layer ($x \approx 0.8$) was sputtered at room temperature. Electron-beam lithography and etching in an SF$_6$ plasma were used to define nanowire meanders. An antireflection coating was deposited on the top surface. A keyhole shape was etched through the Si wafer around each SNSPD, which could then be removed from the wafer and self-aligned to a single mode optical fibre~\cite{Miller2011}. The size of the SNSPD is $16 \times 16$ $ \mu$m$^2$, larger than the 10 $\mu$m mode field diameter of a standard single mode fibre, to accommodate any minor misalignment. The optimal system detection efficiency reaches 70\% with a dark count rate of the order of a few hundred counts per second.

\subsection{ Data acquisition and analysis.} 
Experimentally, the probability of having a qubit $P_\text{in}$ is measured by dividing the number of coincidences between detectors $D_1$-$D_2$ by the number of counts in detector $D_1$. The output probability $P_\text{out} = G(t) P_\text{in}$ is given by dividing the four-fold coincidence $D_1$-$D_2$-$D_3$-$D_4$, by $D_1$-$D_2$-$D_3$. This allows us to determine the gain of the amplifier as a function of  $P_\text{in}$. 
 The coincidence rate is around 10 counts/minute for each of the two curves, $t=0.7$ and $t=0.9$, and it drops linearly with the input probability $P_\text{in}$. For this reason, when measuring the gain, we increase the time of measurement by a factor $1/P_\text{in}$ in order to have the same statistics for each run. The measurement corresponding to the  highest value of $P_\text{in}$ was taken by recording timestamps of the detections for 900 s.
Error bars are calculated assuming poissonian statistics of photon counting.

\section{Result}

\begin{figure}
\includegraphics[width=0.85\columnwidth]{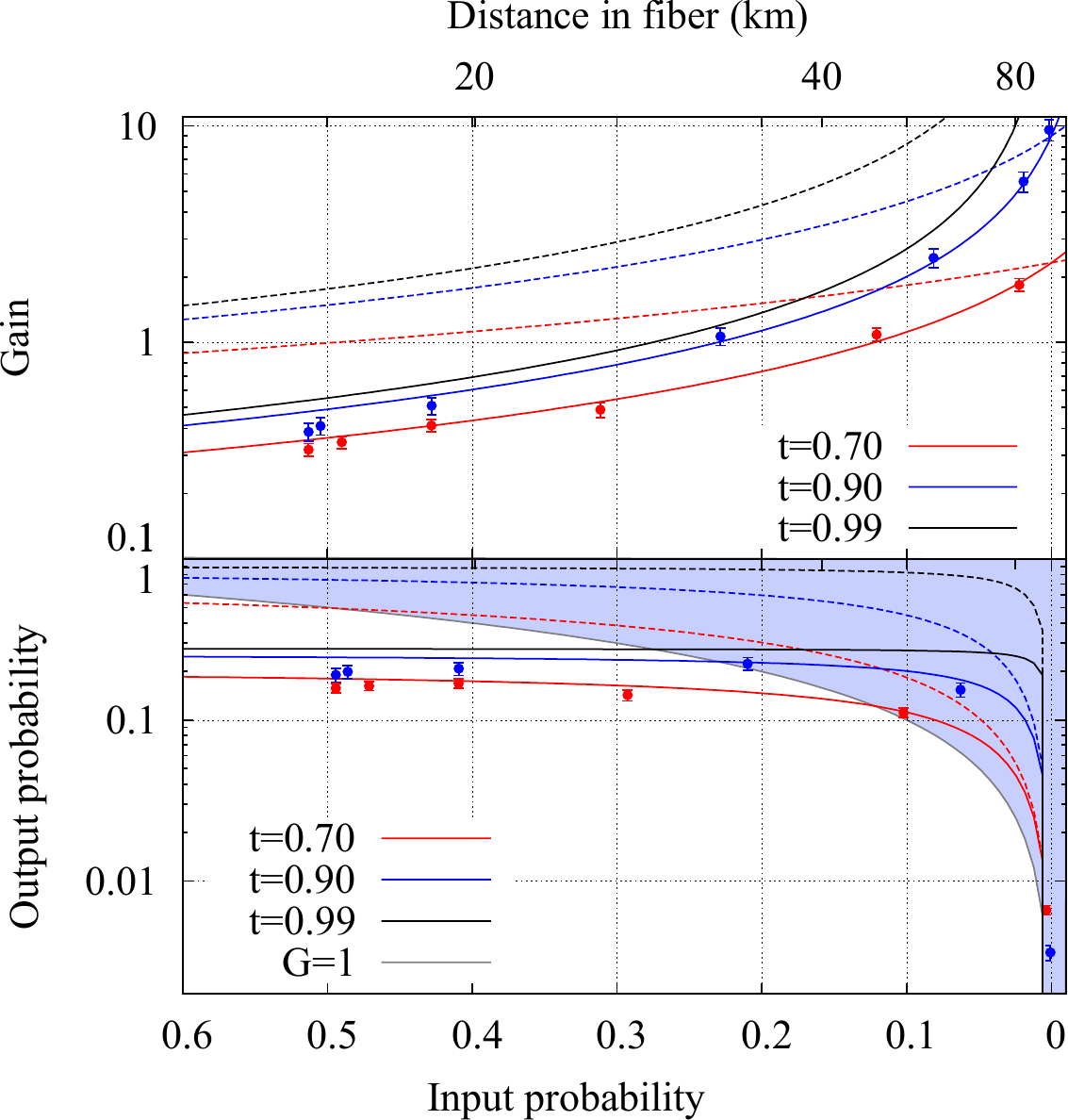}
\caption{Top: Gain as a function of the input probability $P_\text{in}$. The red and blue points correspond to unbalanced beam solitters with transmission $t=0.7$ and $t=0.9$, respectively. Black corresponds to $t=0.99$. Solid lines take in account the measured transmission loss for the ancillae, dashed lines are calculated assuming the best case scenario in which the probability of having an ancilla is $P_\text{a}=0.9$. Bottom: Output probability as a function of the input probability $P_\text{in}$. The colored area indicates where the gain becomes higher than 1, i. e. one actually has amplified the input state. }\label{fig3}
\end{figure}

\begin{figure}
 \includegraphics[width=0.9\columnwidth]{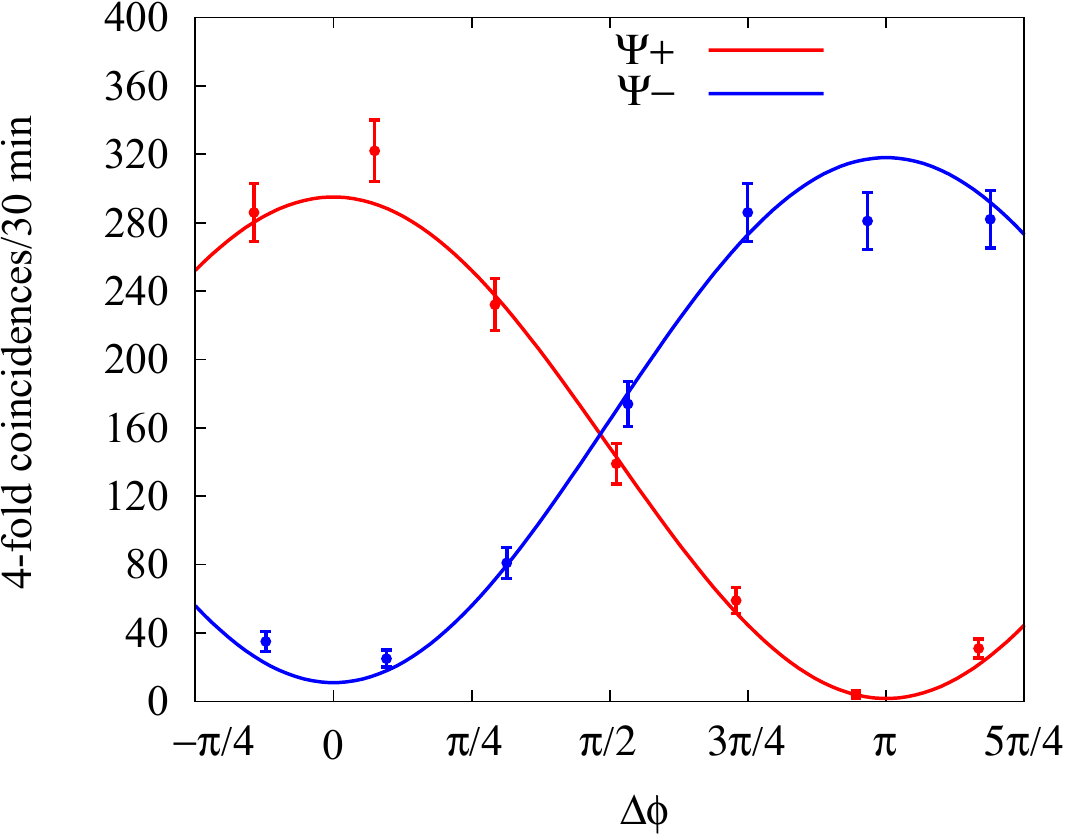}
\caption{Analysis of the output state as a function of $\Delta \phi$. The visibilities are  $0.98 \pm 0.02$ and $0.93 \pm 0.02$ for the red and the blue curve, respectively, leading to fidelities of $0.99 \pm 0.01$  and $0.97 \pm 0.01$ . For this measurement $t=0.7$ and $P_{in}=0.47\pm0.04$.}\label{fig4}
\end{figure}

\figurename{~\ref{fig3}} shows that the experimental results are in good agreement with the theoretical prediction.
The measurement is done for two different values of \textit{t}: 0.7 and 0.9. 
In the case of $t = 0.9$, the gain reaches the maximum value of 9, fixed by $t/(1-t)$.  The dashed lines in \figurename{~\ref{fig3}} show the gain when the excess loss for the ancillae are factored out -- we keep experimentally feasible values for $P_\text{a} = 90\,\%$~\cite{Guerreiro2013}.  
To measure the Fidelity of the output state, defined as $\mathcal{F} = \langle \psi_\text{in} \vert \rho_\text{out} \vert \psi_\text{in} \rangle$, we project the output state on the input. Varying the phase $\Delta \phi$ of the input qubit, we can observe oscillations in the four fold coincidence rates, as shown in \figurename{~\ref{fig4}}. The two curves correspond to the two possible outputs of the BSM: at $\Delta \phi= 0$, the output obtained when projecting onto $\Psi^+$ gives the maximum overlap, while the output related to $\Psi^-$ is orthogonal to the qubit, since a phase shift of $\pi$ (the unitary operation from equation~\ref{eq:rhoout}) is needed to recover the input state. By measuring the visibility of the fringes observed as a function of the phase difference $\Delta \phi$, we can estimate the Fidelity as $\mathcal{F}=(1+V)/2$. . The visibilities for the two curves shown in \figurename{~\ref{fig4}} are $0.98\pm0.02$ and $0.93\pm0.02$, leading to fidelities of $0.99\pm0.01$  and $ 0.97\pm0.01$. 
Each point is measured fixing the phase at a maximum for one output of the BSM and minimum for the other one. 
We use these two points to check that the fidelity is constant for each value of gain.

The main figure of merit is, however, the output probability  $P_\text{out}$, plotted in~\figurename{~\ref{fig3}} as a function of $P_\text{in}$. To be more specific, $P_\text{out}$ is the probability of having a qubit at the output once a successful BSM is announced. From~\figurename{~\ref{fig3}} one can see that the main limitation on the performance of the amplifier is given by the coupling efficiency of the ancilla $P_\text{a}$, which, multiplied by \textit{t}, gives the upper bound on the output probability. Again, the theory and experiment are in good agreement and when the excess loss (ancillae preparation) is factored out we are clearly in a regime of net gain. n.b. if a beam splitting ration of $t = 0.99$ is used, an output probability $P_\text{out} > 82.3\,\%$ is achievable, which would meet the requirements for DIQKD even for distances up to 40\,km.

\begin{figure}
\includegraphics[width=0.85\columnwidth]{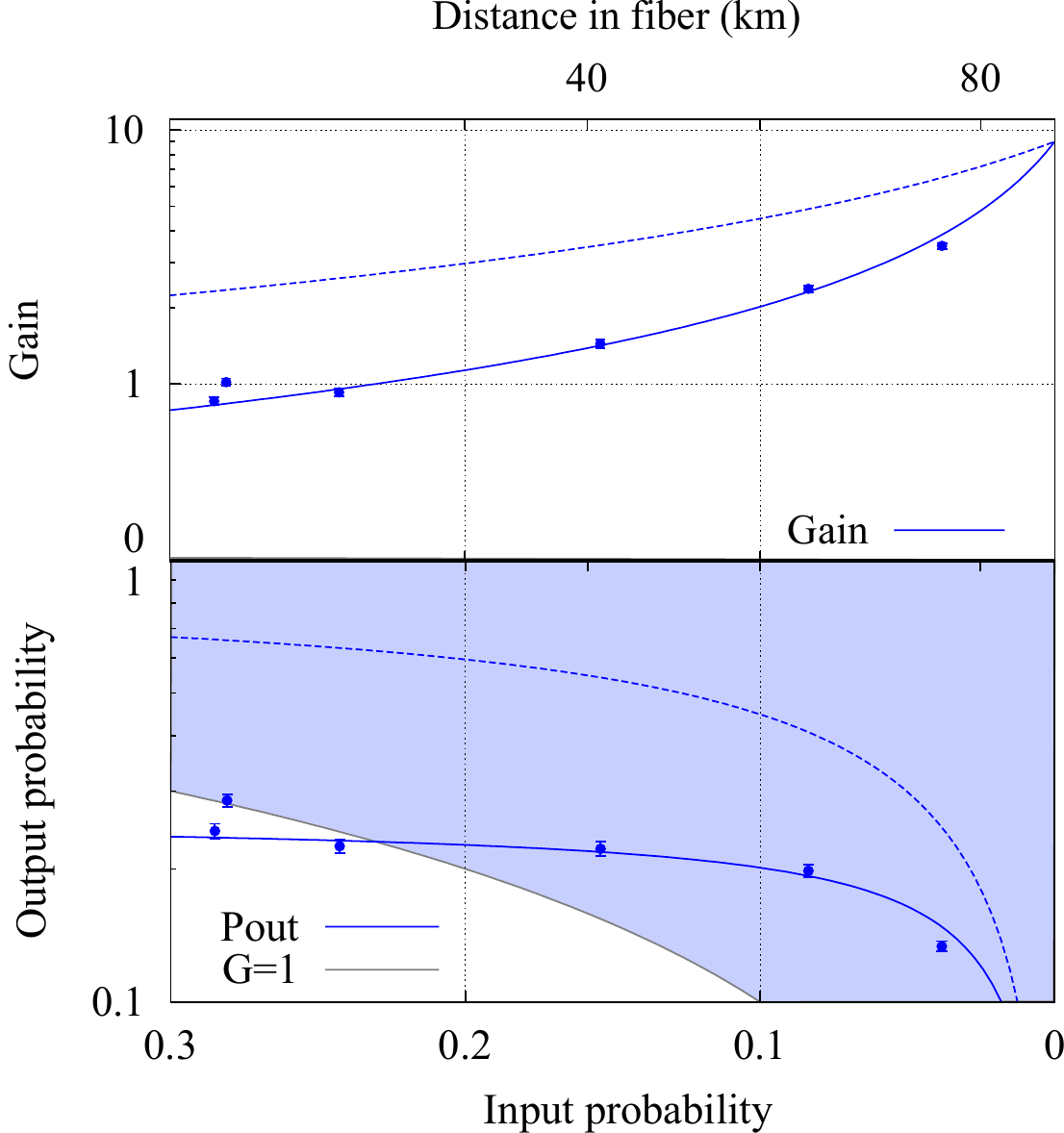}
\caption{ The gain with $t = 0.9$ and output probability, of the single photon heralded qubit amplifier. Solid lines are calculated taking in account the measured transmission losses, dashed lines are calculated assuming $P_\text{a}=0.9$.}\label{fig5}
\end{figure}

Finally, using the same setup, we demonstrate heralded (single) photon amplification (HPA) with no postselection, as proposed in the seminal work of \cite{T.C.Ralph2009}.
 Indeed, this experiment requires two independent heralded single photons, which can be realised with our four-photon source.
In previous works~\cite{Osorio2012,Bruno2013}, some of the authors have shown the heralded amplification of photons using two photons coming from the same SPDC process as input and ancilla. In the present work, referring to the experimental scheme depicted in \figurename{~\ref{fig2}}, the time-bin qubit preparation (and analysis) stage can be removed and only one (heralded) ancilla photon is sent to the amplifier. This enables us to perform non-postselected HPA, i.~e. the amplification is conditioned only upon the detection of the two heralding photons plus the BSM. \figurename{~\ref{fig5}} shows the gain and the output probability as a function of the input probability measured in our setup. As before the dashed lines indicate the performance without excess loss. 

\section{Conclusion}

We have demonstrated an all-fibre heralded qubit amplifier operating in the telecom regime for both time-bin and Fock-state qubits, showing the first post-selection free operation in the case of the latter. The implementation highlights the suitability for a fully integrated photonics solution, where the addition of photonic sources and even detectors could further assist the efficiency of these devices. While a gain of 9 is realised with the current system, the key parameter is the efficiency of the HQA, which is limited primarily by $P_\text{a}$.

Recently, a new scheme for characterizing entanglement for Fock-state systems~\cite{Monteiro2015,Vivoli:2015aa} was demonstrated; by extending this to systems with independent photon sources, as required here, heralded qubit amplification should provide an experimentally feasible path towards loophole-free Bell tests and DIQKD over tens of kilometers.

\section*{Acknowledgments}

The authors would like to thank Alexey Tiranov for discussions. This work was supported by the Swiss NCCR QSIT, the EU projects SIQS and DIQIP and the DARPA Quiness program and the SNSF (Grant No. PP00P2 150579). 
Part of the research was carried out at the Jet Propulsion Laboratory,
California Institute of Technology, under a contract with the National
Aeronautics and Space Administration.

%

\end{document}